
\input phyzzx

\nopubblock
\PHYSREV
\parindent=0 truecm
\parskip=0 truecm

{\bf
\titlepage
\title{DARK MATTER AND BIG BANG NUCLEOSYNTHESIS}
\author{Arnon Dar\foot{Supported in part by the Technion Fund for
Promotion of Research}}
\address{Department of Physics and Space Research Institute,
Technion - Israel Institute of Technology, Haifa 32000, Israel}}
\baselineskip=12pt
\abstract
The recently observed Deuterium abundance in a low-metallicity
high-redshift hydrogen cloud, which is about ten times larger than that
observed in the near interstellar medium, is that
expected from the Standard Big Bang Nucleosynthesis
theory and the observed abundances of $^4$He and $^7$Li
extrapolated to their primordial values. The inferred cosmic
baryon to photon ratio, $\eta=(1.60\pm 0.1)\times 10^{-10},$
yields a mean cosmic baryon density, in critical mass units,
of $\Omega_b\approx (0.6\pm 0.1)\times 10^{-2}h^{-2},$
where h is the Hubble constant in units of $100~km~s^{-1}~Mpc^{-1}$.
This baryon density is consistent with the mean cosmic density of
matter visible optically and in X-rays. It implies that most
of the baryons in the Universe are visible and are not dark.
Combined with the observed ratio of baryons to light in X-ray
emitting clusters, it yields the value $\Omega \approx 0.15$ for the
mean mass density of the Universe,
which is consistent with that obtained from the
mass to light ratio in clusters.
This mass density is about ten times larger than the mean baryon mass
density. It indicates that most of the matter in the Universe consists
of nonbaryonic dark matter.
\bigskip
\endpage
\baselineskip=12pt
{\bf I. INTRODUCTION} The agreement between the predictions of the
Standard Big Bang Nucleosynthesis (SBBN) theory (Peebles 1966;
Wagoner, Fowler and Hoyle 1967; Wagoner 1973; Yang et al 1984)
and the observed
abundances of H, D, $^3$He, $^4$He, and $^7$Li extrapolated to their
primordial values which span about 10 orders of magnitude
is one of the most convincing pieces of supportive evidence for
the Standard Hot Big Bang Model of the early Universe (e.g.,
Weinberg 1972; Kolb and Turner 1990; Peebles 1993).
The predictions of the SBBN theory depend on
low energy nuclear cross sections and on three additional parameters,
the number of flavours of light neutrinos, $ N_\nu~,$
the neutron lifetime, $\tau_n~$, and the ratio of baryons to photons
in the Universe, $\eta\equiv n_b/n_\gamma~$.
The relevant nuclear cross sections are known from laboratory
measurements (e.g., Caughlan and Fowler 1988 and references therein,
Smith et al 1993 and references therein).
Measurements at the Large Electron Positron Collider (LEP) at CERN gave
$ N_\nu=3.04\pm 0.04$ (e.g., Mana and Martinez 1993).
Measurements of $\tau_n$ in neutron bottles
and Penning traps coupled with previous measurements yielded
the weighted average  (see Particle Data Group 1994)
$\tau_n=887\pm 2.0~s~. $ Finally, measurements
of the cosmic microwave background radiation by COBE (Mather et al 1994)
gave a blackbody temperature  $T=2.726\pm 0.017~K,$ which yields
$n_\gamma=20.28T^3\approx 411\pm 8~cm^{-3}. $ Hence,
the primordial abundances of the light elements
are now predicted quite accurately by the SBBN theory
as a function of a single unknown parameter,
$n_b~,$ the mean baryon number density in the Universe.
The primordial abundances of $^4$He, D, $^3$He
and $^7$Li that are predicted by the SBBN theory
with $N_\nu=3$, $\tau_n=887~s$ and the nuclear cross sections that
were compiled by Caughlan and Fowler (1988) and updated by Smith et al
(1993) are displayed in Fig. 1  for $1\leq\eta_{10}\leq 10~.$
Thus, the
primordial abundances of the light elements  which are inferred
from observations, can be used to test
the consistency of the  hot Big Bang model of the early Universe and
to determine the mean baryon density in the Universe.

Indeed, during the past few years it has been claimed  repeatedly
that the predictions of SBBN theory agree with observations
if $\eta_{10}\equiv \eta\times 10^{10}\approx 4$, and that
implies that most of the nucleons in the Universe are dark
(e.g., Kolb and Turner 1990, Walker et al 1991,
Smith et al 1993 and references therein).
Moreover, based on these analyses, variety of
limits on physics beyond the standard particle physics model
(new interactions; new weakly interacting
particles; additional neutrino flavours; masses, mixings,
magnetic moments, decay modes and lifetimes of neutrinos) were
derived by various authors.
\smallskip
However, the claimed concordance between SBBN theory
and the observed abundances of the light elements
extrapolated to their primordial values
had a rather poor confidence level,
was demonstrated for a primordial abundance of $^4$He
that deviated significantly from its best value inferred
from observations, and relied heavily on the
highly uncertain extrapolated values for the
primordial abundances of D+$^3$He. Hence, SBBN
could provide neither reliable evidence that most
of the baryons in the universe are dark nor reliable
limits on the physics beyond the standard particle physics model
(Dar, Goldberg and Rudzsky 1992).

\smallskip
During the past three years new observations and refined analyses
have greatly improved the estimated values of the primordial
abundances of $^4$He
(Pagel et al 1992; Mathews et al 1993;
Skillman and Kennicutt 1993; Izotov et al 1994),
$^7$Li (Thorburn 1994) and
D (Songaila et al 1994; Carswell et al 1994). In particular, the
recent measurements of the absorption  spectrum of the distant  quasar
Q0014+813 in a low-metallicity high-redshift (z= 3.32) intervening
hydrogen cloud, by Songaila et al
(1994) with the Keck 10m telescope at Mauna Kea, Hawaii, and
by Carswell et al (1994) with the 4m telescope at Kit Peak, Arizona
show an absorption line
whose position coincides with that expected
of the isotopically shifted Lyman
$\alpha$ absorption line of deuterium. The
best fitted  Deuterium abundance
is ${\rm 1.9\times 10^{-4}\le [D]/[H] \le 2.5\times 10^{-4}}$.
Below I show (see also Dar 1995a; 1995b)
that the above value for the primordial
abundance of Deuterium  and the best estimated
values of the primordial abundances
of $^4$He and $^7$Li which were inferred from observations are
in excellent agreement with those predicted by SBBN theory,
if the cosmic baryon to photon
ratio is $\eta=(1.60\pm 0.1)\times 10^{-10}.$
This ratio yields a mean cosmic baryon density, in critical mass units,
of $\Omega_b\approx (0.6\pm 0.1)\times 10^{-2}h^{-2},$
with $h$ being the Hubble constant in units of $100~km~s^{-1}~Mpc^{-1}$,
which is consistent with the mean baryon density of matter visible
optically and in X-rays. It implies that (a) most
of the baryons in the Universe are visible and are not dark, (b)
$\Omega \approx 0.15$ as inferred also from the mean luminosity
density of the Universe and the masses
of clusters of galaxies obtained from optical, gravitational lensing
and X-ray observations, (c) most of
the dark matter in groups and clusters of galaxies is non baryonic,
(d) most of the matter in the Universe is nonbaryonic dark matter.

\bigskip
{\bf III. INFERRED PRIMORDIAL ABUNDANCES}
\smallskip
{\bf Helium 4:}
The most accurate determinations
of the primordial abundance of $^4$He are based on measurements of
its recombination radiation in very low metallicity extragalactic
HII regions which are the least contaminated by stellar production of
$^4$He. A number of groups have obtained high-quality data for very
metal-poor, extragalactic HII regions which they
used to extrapolate to zero metallicity yielding
$Y_p=0.228\pm 0.005$ (Pagel et al
1992), $Y_p=0.226\pm 0.005$ (Mathews et al 1993),
$Y_p=0.230\pm 0.005$ (Skillman and Kennicutt 1993),
$Y_p=0.232\pm 0.004$ (Izotov et al 1994),
where $1\sigma$ statistical
and systematic errors were added in quadrature. A weighted
average yields
$$Y_p=0.229\pm 0.005~. \eqno\eq $$
It is not inconceivable that systematic errors (e.g., due to collisional
excitation, contribution of neutral Helium, interstellar reddening,
UV ionizing radiation, grain depletion, nonhomogeneous density and
temperature, etc.) are larger.
However, there is no empirical indication for that.

{\bf Deuterium:}
Since Deuterium is easily destroyed at relatively low temperatures.
its abundance observed today can only
provide a lower limit to the big-bang production.
Measurements of the Deuterium abundance
in the local interstellar medium (LISM)
made recently by the Hubble Space Telescope (Linsky et al. 1993), gave
 [D]/[H]$ = (1.65^{+0.07}_{-0.18})\times 10^{-5}.$
{}From the analysis of solar-wind particles captured
in foils exposed on the moon and studies of primitive meteorites,
Geiss (1993) deduced a pre-solar Deuterium abundance
[D]/[H]$ = (2.6 \pm 1.0)\times 10^{-5}.$  These values can be used as
lower bounds on primordial Deuterium.
High-redshift, low-metallicity quasar absorption systems
offer the possibility of observing Deuterium
abundance back in the past in very primitive clouds (Webb 1991).
Recently, Songaila et al (1994) and Carswell et (1994)
detected in a high-redshift, low-metallicity  cloud
an absorption system at Z=3.32 in the spectrum of the
Quasar Q0014+813 with an absorption line at the expected position of the
isotopically shifted Ly$\alpha$ line of Deuterium. From fitting
the line shape they found a Deuterium abundance of
$${\rm [D]/[H] = (1.9 - 2.5) \times 10^{-4}}. \eqno\eq $$
The probability that the absorption line is due to a second intervening
small hydrogen cloud with the Ly$\alpha$ absorption line
at the position of the isotopically shifted deuterium line,
(this requires a relative velocity of 82 $km~s^{-1}$)
was estimated as 3\% and 15\% by Songaila et al (1994)
and by Carswell et al (1994), respectively.
Preliminary results by Tytler et al, Carswell et al, and Cowie
et al on the [D/H] ratio in absorption systems of other quasars
yield a range of values between $2\times 10^{-5}$ and
$2\times 10^{-4}$. The primordial [D/H] ratio, however, should be
obtained by extrapolating the measured values in
the most metal poor absorbers to zero metallicity.
The above value of
$2\times 10^{-4}$ is an order of magnitude larger than the
interstellar value and a factor of three larger than the 95\%
confidence level upper bound on the primordial abundance
of D+$^3$He that was inferred by Walker et al (1991).
However, Walker et al (1991) used an uncertain galactic chemical
evolution model to extrapolate their estimated
presolar D+$^3$He abundance to zero cosmic age.
Moreover, measurements of D and $^3$He abundances in the interstellar
space within the Milky Way
show large variations from site to site and the solar system values
may not be a typical sample of galactic material 4.5 Gyr ago.

{\bf Helium 3:} From measurements of $[^3$He]/$[^4$He] in meteorites
and the solar wind Geiss (1993) concluded that the presolar
abundance of $^3$He is $[^3$He]/[H]=$(1.5\pm 0.3)\times 10^{-5}$.
However, any further extrapolations to zero cosmic age
of the $^3$He abundance extracted
from solar system or interstellar observations are highly
uncertain because
$^3$He is both produced [via D(p,$\gamma)^3$He] and destroyed
[via $^3$He($^3$He,2p)$^4$He and $^4$He($^3$He,$\gamma)^7$Be]
in early generation stars.
(Hogan 1994 has suggested recently that
the envelope material in low mass stars is mixed
down to high temperature after they reach the giant branch, so that the
$^3$He is destroyed before the material is ejected.)
Indeed, from radio observations
of highly ionized Galactic HII regions Balser et al (1994)
and Wilson and Rood (1994) inferred  [$^3$He]/[H] values that
ranged between
 $(6.8\pm 1.5)\times 10^{-6}$ for W49 and
 $(4.22\pm 0.08)\times 10^{-5}$ for W3.
Hyperfine emission in the  planetary nebula N3242
indicates  (Rood, Bania and Wilson 1992)
a large enrichment [$^3$He]/[H]$\approx 10^{-3}$.
These values show
that the presently observed $^3$He abundances apparently reflect
complicated local chemical evolution and do not allow
a reliable determination of the primordial abundances of either
$^3$He or $^3$He+D from
presently observed solar or LISM abundances.

{\bf Lithium 7:} The primordial abundance of $^7$Li ,
was determined from the most metal poor, Population II halo stars.
Such stars, if sufficiently warm $(T\gsim 5500K)$,
have apparently not depleted their surface Lithium and are expected
to have a nearly constant Lithium abundance reflecting the Lithium
abundance present at the early evolution of the Galaxy (Spite and
Spite 1982a, b).
High-precision LiI observations of 90 extremely metal
poor halo dwarfs and main sequence turnoff stars
have been performed recently by Thorburn (1994).
{}From the surface lithium abundances of the hottest metal-deficient
stars ($ T\sim 6400K)$ Thorburn estimated
$${\rm[^7Li]_p/[H]_p=(1.7\pm 0.4)\times 10^{-10}}.\eqno\eq $$
Thorburn's data suggest a slight
systematic variation of the $^7$Li abundance with surface
temperature, possibly indicating some depletion from a higher
primordial value by processes that transport $^7$Li inward to
regions where it can be burned.
However, the amount of depletion
is constrained by the relatively narrow spread in $^7$Li abundance
for a wide range of surface temperatures and metallicities and
by the observation of $^6$Li in population
II stars by Smith, Lambert, and Nissen (1993) and by
Thorburn (1994):
Big-bang production of $^6$Li is negligible. It is
presumably produced by cosmic-rays.
Since $^6$Li is burned much more easily than
$^7$Li and yet still observed with the abundance
expected for cosmic-ray production, depletion of $^7$Li
cannot have been very significant.
\bigskip
{\bf IV. COMPARISON BETWEEN THEORY AND OBSERVATIONS}
\smallskip
In Fig 1 we compare the predictions of the
SBBN theory and the observed abundances of the light elements
extrapolated to their primordial values (summarized in section III).
The confidence level of the agreement between the two
using the standard $\chi^2$ test as function of
$\eta_{10}$ is also shown in Fig.1.
(Errors were assumed to be statistical in nature. Experimental and
theoretical uncertainties were added in quadrature). Fig.1 shows
that the primordial abundances of the light elements as inferred
from observations are in very good agreement
(confidence level higher than 70\%) with those predicted by SBBN theory
provided that $\eta_{10}\approx 1.60\pm 0.1~.$
The corresponding mean cosmic baryon number density is
$ n_b=\eta n_{\gamma}= (6.6\pm 0.5)\times
10^{- 8}cm^{-3},$  which yields a baryon mass density
(in critical density units $\rho_c\equiv 3H_0^2/8\pi G~$) of
$$\Omega_b\equiv \rho_b/\rho_c=(0.6\pm 0.1)\times10^{-2}
h^{-2}.\eqno\eq $$
and SBBN abundances of
$Y_p=0.230\pm 0.002$,
${\rm [D]_p/[H]_p= (2.12\pm 0.20)\times 10^{-4}}$,
${\rm [^3He]_p/[H]_p=(2.38\pm 0.08)\times 10^{-5}}~$ and
${\rm [^7Li]_p/[H]_p=(1.88\pm 0.44)\times 10^{-10}}~.$
\bigskip
{\bf V. ARE MOST BARYONS DARK ?}
\smallskip
Baryons are visible when they form stars or when they emit or absorb
electromagnetic radiation in neutral and ionized gas.

Most of the visible stars are within the optical radius of
galaxies. The mean numbers of galaxies per unit volume with luminosity
in the range L to L+dL is well represented by the Schechter luminosity
function (Schechter 1976):
$$ dn=\phi(L) dL=\phi_*(L/ L_*)^\alpha
       e^{-L/L_*}(dL/L_*)~. \eqno\eq $$
where $\phi_*$ is a normalization constant, $\alpha$ is a power
parameter and $L_*$ is the luminosity of a typical galaxy.
Recent measurements (Loveday et al 1992) gave
$\phi_*=(1.40\pm 0.17)\times 10^{-2}L_\odot~h Mpc^{-3},$
$L_*=(1.21\pm  0.15 )\times 10^{10}L_\odot $ and
$\alpha=-1.11\pm 0.15~.$
About 1/3 of the galaxies are ellipticals and 2/3 are spirals.
Within their optical radii, the ratio of mass to blue light
of ellipticals can be represented approximately by
$(M/L_B)=R(M_\odot / L_\odot)(L/ L_*)^\beta~$
with (e.g., van der Marel 1991) $\beta\approx 0.35\pm 0.05.$
Consequently for ellipticals, within their optical radius,
$<M/L_B>=(8\pm 2)h~.$  Similarly,
for spirals $<M/L_B>\approx (2.4\pm 0.6)h$ within their optical radius.
Hence, the mean cosmic densities of light and mass
within the luminous part of galaxies are given, respectively, by
$$\rho_L=\phi_* \Gamma(2+\alpha)L_*=(1.83\pm 0.35)\times 10^8h
L_\odot~Mpc^{-3}, \eqno\eq $$
$$\rho_*=\phi_* \Gamma(2+\alpha+\beta)R M_\odot L_*/L_\odot
=(1.23\pm 0.38)\times 10^9h^2M_\odot~Mpc^{-3},\eqno\eq $$
yielding
$\Omega_* = (0.28\pm 0.10)\times 10^{-2}$ (similar to the value
derived by Persic and Salucci (1992).
Note, however, that the luminous part of galaxies may contain
as much as $50\%$ dark matter which may be nonbaryonic, i.e.,
the mean density of baryons in stars may be as small as
$\Omega_* = (0.14\pm 0.05)\times 10^{-2}.$
\smallskip
Recent X-ray observations indicate that clusters and groups of
galaxies contain a total mass of intergalactic gas
much larger than their total galactic mass (e.g., White et al 1993
and references therein).
For instance, whereas in  the Milky Way galaxy
the ratio of total gas mass to light is
$M_{gas}/L_{MW}\approx 4.8\times 10^9M_\odot/ 2.3\times 10^{10}L_\odot
\approx 0.21 M_\odot/L_\odot$,
analyses of recent observations
with the ROSAT X-ray telescope  of the compact group HCG62 and the Coma
cluster yielded $M_{gas}/L_B\approx 4.4\times 10^{11}M_\odot
h^{-5/2}/2.4\times 10^{10}h^{-2}L_\odot
\approx 19h^{-1/2} M_\odot/L_\odot$ within a distance of
$0.24h^{-1}~Mpc$ from the center of HCG62 (Ponman et al 1993), and
$M_{gas}/L_B\approx (5.45\pm 0.98) \times 10^{13}M_\odot
h^{-5/2}/ 1.95\times 10^{12}h^{-2}L_\odot
\approx (28\pm 6) h^{-1/2} M_\odot/L_\odot$
within a distance of $1.5h^{-1}~Mpc$ from the center
of the Coma cluster (Briel et al 1992; White et al 1993).
These ratios are approximately radius
independent because the
optical measurements of the velocity dispersion of galaxies
in clusters and the X-ray measurements of the temperature profile
of the X-ray emitting intracluster gas show that both the galaxies
and the intergalactic gas in clusters trace the same potential and
have similar velocities and density distributions.
If $<M_{gas}/L_B>\approx (23\pm 5)h^{-1/2}$ for clusters and
groups represents well this ratio for the whole Universe, then
from the measured density of blue light
in the Universe  (see Eq. (6))
we obtain, respectively, the mean cosmic densities of gas and baryons,
$$\Omega_{gas}={\rho_L<M_{gas}/L_B>\over \rho_c}
\approx (1.52\pm 0.45)\times 10^{-2}h^{1/2}, \eqno\eq $$
$$\Omega_b=\Omega_{gas}+{\rho_L<M_*/L_B>\over \rho_c}
\approx ([1.52\pm 0.45]h^{1/2}+0.28\pm 0.15)\times 10^{-2}.\eqno\eq $$
Recent measurements of the Hubble parameter
(e.g. Freedman et al 1994; Pierce et al 1994; Dahle et al 1994)
indicate that  $h=0.70\pm 0.10$, yielding $\Omega_b=
(1.5\pm 0.5)\times 10^{-2}.$
This estimated mean  density of baryons which
are visible optically and in X-rays,
is consistent with the mean density of baryons
obtained from BBNS, $\Omega_b=(1.2^{+0.5}_{-0.3}\times 10^{-2}$.
Therefore we conclude
that BBNS does not provide evidence that most of the baryons are dark.

The argument that a large fraction of baryons are dark
(e.g., Kolb and Turner
1991; Walker et al 1991) was based on the extrapolations of the
observed abundances of Deuterium in the solar system (e.g., Geiss 1993)
and in  the local interstellar medium (Linsky et al 1993) to zero age.
Its validity requires not only that the primordial
abundance of D which was determined from the absorption spectrum
of the quasar Q0014+813 is wrong, but also
that the primordial abundances of $^4$He  and $^7$Li are much
larger than their current best estimated values. Only then
a large fraction of the total baryonic mass can be dark. Such baryonic
dark matter could reside in low mass stars and brown dwarfs, which can
produce microlensing events like those recently discovered by the MACHO
(Alcock et al 1993; Sutherland 1995),
EROS (Aubourg et al 1993; Moscoso 1995) and OGLE (Udalski et
al 1993) collaborations. However, recent measurements with the Hubble
Space Telescopes indicate that in the Milky Way
very little baryonic mass resides in low mass stars, i.e., that the
mass function of stars does not increase suddenly towards low
stellar masses (Bahcall et al 1994). Moreover, preliminary analysis
of the statistics of microlensing events yields an optical
depth for microlensing of LMC stars by dark massive halo objects
(MHO) much smaller than that expected if the
MW mass within a radius equal to the distance to the LMC
was made of MHOs (Alcock et al 1995).

\smallskip
{\bf VI. Evidence For Non Baryonic Dark Matter}
\smallskip
Rich clusters of galaxies are the largest objects for which
total masses can be estimated directly. In fact, the need for
astrophysical dark matter was first identified for such
systems by Zwicky in 1933.

The total mass enclosed within a distance R from the centers
of clusters of galaxies has been determined by three independent
methods:

a) From the virial theorem applied to the radial velocities
of cluster members assuming that the velocities are
distributed isotropically and that light traces mass (see e.g.,
Peebles 1993).

b) From analyses of the distribution of giant arcs and arclets
produced by gravitational lensing of distant galaxies
by the gravitating mass in clusters of galaxies.
(see e.g., Tyson et al 1990; Tyson 1991; Kaiser and Squires 1993;
Smail et al 1994; Soucail, this volume).

c) From the X-ray emission of intergalactic hot gas which is
trapped in the deep gravitational potential of rich clusters,
under the assumption that the gas is relaxed (see for instance
Jones and Forman 1984; Sarazin 1986;
Bohringer, this volume; Forman, this volume).
\smallskip
All three methods yield similar results. When coupled with
photometric measurements of the light emitted by the galaxies
in the clusters they yield (see, e.g., Bohringer, this volume;
Forman, this volume) an average total mass to blue light ratio
of $<M/L_B>= (230\pm 30)hM_\odot/L_\odot.$ The density of blue light
in the Universe was measured (e.g., Loveday et al 1992)
to be  $\rho_L=(1.83 \pm 0.35)\times 10^8h~Mpc^{-3}$.
If the mean M/L ratio for clusters  represents well the mean M/L
ratio in the Universe then the mean cosmic density is
$$\Omega={\rho_L<M/L>\over \rho_c}\approx 0.15\pm 0.3~.
\eqno\eq $$
This mean cosmic density is larger
by more than an order of magnitude than the mean baryon density inferred
from SBBN. This difference provides the best observational
evidence for non baryonic dark matter. It is also
confirmed by the recent observations with the ROSAT  X-ray telescope
which has been used to
determine the total gravitating mass, $M_t$, of clusters
and the fraction of that mass which is in the form of X-ray
emitting gas, $M_{gas}$, and by photometric measurements of the light
emitted by the galaxies in the clusters which have been used to estimate
the total ``stellar''
mass, $M_*$, within the visible part of these galaxies.
It was found (e.g.,
Briel et al 1992; White et al 1993; White and Fabian 1995; Buote
and Canizares 1995) that
$M_*/M_t\approx 0.01$  and  $<M_{gas}/M_t>\approx 0.05h^{-3/2}$,
i.e., the known forms of baryonic matter account only for
a small fraction of the total mass of x-ray emitting clusters.
If clusters constitute a fair sample of the Universe it implies
that most of the matter in the Universe is non baryonic dark matter.
\smallskip
In fact, numerical simulations
of structure formation indicate that the ratio of baryonic to non
baryonic mass is preserved in cluster formation (e.g., White et al
1993). Consequently, the mean baryonic density deduced from SBBN
and the observed baryonic fraction in clusters imply that
$$ \Omega\approx {M_t\over M_b}\Omega_b\approx {0.0058h^{-2}\over
   0.01+0.05h^{-3/2}}\approx 0.12\pm 0.2~, \eqno\eq $$
in good agreement with the independent estimate from the mass
to light ratio of clusters.

It is interesting to note that
if the cosmic dark matter consists of massive neutrinos then
the  neutrino masses satisfy $(\Omega-\Omega_b)\rho_c
 \approx\Sigma n_\nu m_\nu c^2$, where $n_\nu=(3/11)n_\gamma
\approx 112~ cm^{-3}$ for each neutrino flavour. Consequently,
for $h=0.7\pm 0.1$, one obtains
$\Sigma m_\nu c^2 \approx (12\pm 3)h^2~eV\approx
6\pm 3 ~eV$. Such massive neutrinos, together with the cluster baryons,
can generate in a self consistent way (Tremaine and Gunn 1979)
the gravitational potentials within clusters of galaxies
as determined from the X-ray measurements, the optical measurements and
the gravitational lensing observations.
\bigskip
{\bf VI. CONCLUSIONS}
\smallskip
The  predictions of the SBBN theory and the best values of the
primordial abundances of the light elements extracted from observations
agree very well if the mean cosmic baryon to photon ratio is
$\eta_{10}\approx 1.6\pm 0.1~.$ This value of $\eta$ yields a mean baryon
mass density, $\Omega_b= (0.6\pm 0.1)\times 10^{-2}h^{-2}\approx
(1.25\pm 0.45)~10^{-2},$
consistent with the best estimated value of matter visible
optically and in X-rays $\Omega_L
\approx ([1.52\pm 0.45]h^{1/2}+0.28\pm 0.15)\times 10^{-2}\approx
(1.5\pm 0.5)\times 10^{-2}.$
Thus, SBBN  does not provide evidence
that most of the baryons in the Universe are dark. However,
SBBN and the baryon mass to light ratio in clusters imply
a total mass density of the Universe of $\Omega_b\approx 0.12\pm 0.03$,
consistent with $\Omega\approx 0.15\pm 0.05$, which was
derived from the mean mass to light ratio
of clusters obtained from optical, X-ray and gravitational lensing
observations, and from the observed luminosity density of the Universe.
Since $\Omega \gg \Omega_b~,$ it implies that
most of the matter in the Universe is nonbaryonic
dark matter. If the value of the primordial abundance of Deuterium
that was inferred from the absorption spectrum of the quasar
Q0014+813 is wrong and if its
true value is close to that observed in the local interstellar
medium, and if the primordial abundance of $^4$He is much larger than
its current best estimated value, and if the primordial abundance
of $^7$Li is much larger than its current best estimated value, only then
can a large fraction of the total baryonic mass
reside in low-mass, faint or invisible stars, which can produce
microlensing events like those recently discovered by the MACHO
(Alcock et al 1993; Sutherland 1995),
EROS (Aubourg et al 1993; Moscoso 1995) and OGLE (Udalski et
al 1993) collaborations. However, no excess of low mass stars
was discovered in deep surveys with the Hubble space telescope,
and too few events of microlensed LMC stars were discovered by
MACHO and EROS to support this possibility.
\endpage
\centerline {{\bf REFERENCES}}
\bigskip
Alcock, C. {\it et al.}, 1993, Nature, {\bf 365 }, 621

Alcock, C. {\it et al.}, 1995, PRL, {\bf 74}, 2867

Aubourg, E. {\it et al.}, 1993, Nature, {\bf 365}, 623

Bahcall, J. {\it et al.}, 1994, ApJ. {\bf 435}, L51

Bahcall, N.A., 1988, ARA\&A, {\bf 26}, 631

Bahcall, N.A. and Cen, R.Y., 1993, ApJ. {\bf 407}, L49

Balser, D.S. {\it et al.}, 1994, ApJ. {\bf 430}, 667

Bohringer, H. {\it et al.}, (1994) Nature, {\bf 368}, 828

Buote, D.A. and Canizares, C.R., 1995, ApJ. (submitted)

Briel, U.G. {\it et al.}, (1992), A\&A, {\bf
259}, L31

Carswell, R.F. {\it et al.}, 1994,  MNRAS, in press

Caughlan, G.R. and Fowler, W.A., 1988, Atom. and Nucl. Data Tables,
{\bf 40}, 284

Dahle, H. {\it et al.}, 1994, ApJ. {\bf 435}, L79

Dar, A., 1995a, NP. B (Proc. Suppl.) {\bf 38}, 405

Dar, A., 1995b, ApJ. (in press).

Dar, A., Goldberg, J. and Rudzsky M., 1992, Technion preprint Ph-92-01

Eyles, C.J. {\it et al.}, 1991, ApJ. {\bf 376}, 23

Freedman, W.L. {\it et al.}, 1994, Nature, {\bf 371}, 757

Geiss, J., 1993, in {\it Origin and Evolution of the Elements}, eds.
N. Prantzos, E. Vangioni-Flam, and M. Casse, (Cambridge University Press,
1993) p. 89

Hogan, C., 1994, preprint  astro-ph/9407038

Hughes, J.P., 1989, ApJ. {\bf 337}, 21

Izotov, Y., Thuan. X.T. and Lipovetsky, V.A., 1994, ApJ. {\bf 435}, 647

Jones, C. and Forman, W., 1984, ApJ. {\bf 276}, 38

Kaiser, N. and Squires, G. 1993, ApJ. {\bf 404}, 441.

Kolb, R. and Turner, M., 1990, {\it The Early Universe} (Addison
Wesley-1990)

Linsky, J.L. {\it et al.}, 1993, ApJ. {\bf 402}, 694

Loveday, J. {\it et al.}, 1992, ApJ. {\bf 390}, 338

Mana, C. and Martinez, M., 1993, NP. B (Suppl.) {\bf 31}, 163

Mather, J.C. {\it et al.}, 1994,  ApJ., {\bf 420}, 439

Mathews, G.J. {\it et al.}, 1993, ApJ.,  {\bf 403}, 65

Moscoso, L., 1995, NP. B (Proc. Suppl.) {\bf 38}, 387

Paczynski, B., 1986, ApJ. {\bf 304}, 1

Pagel, E.J. {\it et al.}, 1992, MNRAS, {\bf 255}, 325

Particle Data Group, 1994,  PR, {\bf D50}, 1173

Peebles, P.J.E., 1966, ApJ. {\bf 146}, 542

Peebles, P.J.E., 1993, {\it Principles of Physical Cosmology},
(Princeton Series in Physics, 1993)

Persic, M. and Salucci, P., MNRAS, 1992,
{\bf 258}, 14p

Pierce, M.J. {\it et al.}, 1994, Nature, {\bf 371}, 385

Ponman, T.J. {\it et al.}, 1993, Nature, {\bf 363}, 51

Rood, R.T., Bania, T.M., and   Wilson, T.L. 1992, Nature,
{\bf 355}, 618

Sarazin, C.L., 1986, RMP, {\bf 58}, 1

Schechter, P.L., 1976, ApJ. {\bf 203}, 267

Skillman E.D. and Kennicutt, R.C. Jr., 1993, ApJ. {\bf 411}, 655

Smail, J., Ellis, R. and Fitchett, M., 1994, MNRAS (in press)

Smith, V.V., Lambert, D.L. and Nissen, P.E., 1993, ApJ.
 {\bf 408}, 262

Smith, M.S., Kawano, L.H. and Malaney, R.A., 1993, ApJ. (Suppl.),
{\bf 85}, 219

Songaila, A. {\it et al.}, 1994, Nature, {\bf 368}, 599

Spite, M. and Spite, F., 1982a, A\&A, {\bf 115}, 357

Spite, M. and Spite, F., 1982b, Nature, {\bf 297}, 483

Sutherland, W. {\it et al.}, 1995, NP. B (Proc. Suppl.)
{\bf 38}, 389

Thorburn, J.A., 1994, ApJ. {\bf 421} 318

Tremaine, S. and Gunn, J.E. 1979, PRL, {\bf 42}, 407

Tyson J.A. {\it et al.}, 1990, ApJ. {\bf 349}, L1

Tyson, J.A., 1991, A.I.P. Conf. Proc. {\bf 222}, 437

Udalski, A., {\it et al.}, 1993, Acta Astronomica, {\bf 43}, 289

van der Marel, R.P., MNRAS, 1991, {\bf 253}, 710

Wagoner, R.V., 1973, ApJ. {\bf 179}, 343

Wagoner, R.V., Fowler, W.A. and Hoyle, F., 1967, ApJ. {\bf 148}, 3

Walker, T.P. {\it et al.}, 1991, ApJ. {\bf 376}, 51

Watt, M.P. {\it et al.}, 1992, MNRAS, {\bf 258}, 738

Webb, J.K., 1991,  MNRAS, {\bf 250}, 657

Weinberg, S., 1972 {\it Gravitation And Cosmology} (John Wiley, 1972)

White, S.D.M. {\it et al.}, 1993, Nature {\bf 366}, 429

White, D.A. and Fabian, A.C., 1995, MNRAS, {\bf 269}, 589

Wilson, T.L. and Rood, R.T., 1994, ARA\&A, {\bf 32}, 191

Yang, J. {\it et al.}, 1984, ApJ. {\bf 227}, 697

\bigskip
\centerline{{\bf FIGURE CAPTION}}
\smallskip
{\bf Figure 1.} (a) The primordial mass fraction of $^4$He and
the abundances (by numbers) of D, $^3$He and $^7$Li
as a function of $\eta_{10}$ as predicted by SBBN theory.
Also shown are their observed values extrapolated to zero age,
as summarized in section III.
The vertical line indicates the value $\eta_{10}=1.6$.
(b) The values of $\chi^2$ (left scale)
and the corresponding confidence level (right scale)
of the agreement between the predicted abundances and those
inferred from observations, as function of $\eta_{10}~.$
Best agreement is obtained for $\eta_{10}\approx 1.60~$ with
a confidence level above 70\%.

\end{document}